# Efficient Er/O Doped Silicon Light-Emitting Diodes at Communication Wavelength by Deep Cooling

Huimin Wen[1,#], Jiajing He[1,#], Jin Hong[2,#], Fangyu Yue[2*] & Yaping Dan[1*]

[1]University of Michigan-Shanghai Jiao Tong University Joint Institute, Shanghai Jiao Tong University, Shanghai, 200240, China.

[2]Key Laboratory of Polar Materials and Devices, Ministry of Education, East China Normal University, Shanghai, 200241, China.

[#]These authors contributed equally

* To whom correspondence should be addressed: Yaping.dan@sjtu.edu.cn, fyyue@ee.ecnu.edu.cn

**Abstract**

A silicon light source at communication wavelength is the bottleneck for developing monolithically integrated silicon photonics. Doping silicon with erbium ions was believed to be one of the most promising approaches but suffers from the aggregation of erbium ions that are efficient non-radiative centers, formed during the standard rapid thermal treatment. Here, we apply a deep cooling process following the high-temperature annealing to suppress the aggregation of erbium ions by flushing with Helium gas cooled in liquid nitrogen. The resultant light emitting efficiency is increased to a record 14% at room temperature, two orders of magnitude higher than the sample treated by the standard rapid thermal annealing. The deep-cooling-processed Si samples were further made into light-emitting diodes. Bright electroluminescence with a spectral peak at 1.54 μm from the silicon-based diodes was also observed at room temperature. With these results, it is promising to develop efficient silicon lasers at communication wavelength for the monolithically integrated silicon photonics.

The monolithic integration of fiber optics with complementary metal-oxide-semiconductor (CMOS) integrated circuits will significantly speed up the computing and data transmission of current communication networks.[1-3] This technology requires efficient silicon-based lasers at communication wavelengths.[4, 5] Unfortunately, silicon is an indirect bandgap semiconductor and cannot emit light at communication wavelengths. Erbium ions ($Er^{3+}$) have radiative emissions at 1.54 μm which is in the telecommunication wavelength band for long haul optic fibers.[5-8] Introducing erbium ions (often with oxygen) into silicon by ion implantation was believed to be one of most promising approaches[8-10] among others[11-16] to create silicon-based lasers. However, during the standard rapid thermal treatment, erbium dopants aggregate to form clusters that are efficient nonradiative recombination centers.[17, 18] As a result, the emission from erbium doped silicon is extremely weak at room temperature.[5-10, 17, 18] Here, we applied a deep cooling (DC) process following the high-temperature thermal annealing on the erbium and oxygen co-doped silicon wafer at a cooling rate up to 1000 ℃/s by flushing with Helium gas cooled in liquid nitrogen (77 K). The dramatic cooling process suppresses the aggregation of Er dopants that occurs during the slow cooling process of the high-temperature thermal annealing, allowing for a strong room-temperature emission at 1.54 μm from the Er/O co-doped silicon wafer. The internal quantum efficiency of the emission reaches a record of 14%, approximately 2 orders of magnitude higher than the samples treated by the standard rapid thermal annealing (RTA). The DC-processed Si samples were further made into light-emitting diodes

(LED). Bright electroluminescence with a spectral peak at 1.54 μm from the silicon-based diodes was also observed at room temperature. Given the single crystalline nature of the silicon substrate, efficient silicon lasers may be readily developed for the full integration of fiber optics with CMOS circuitry.

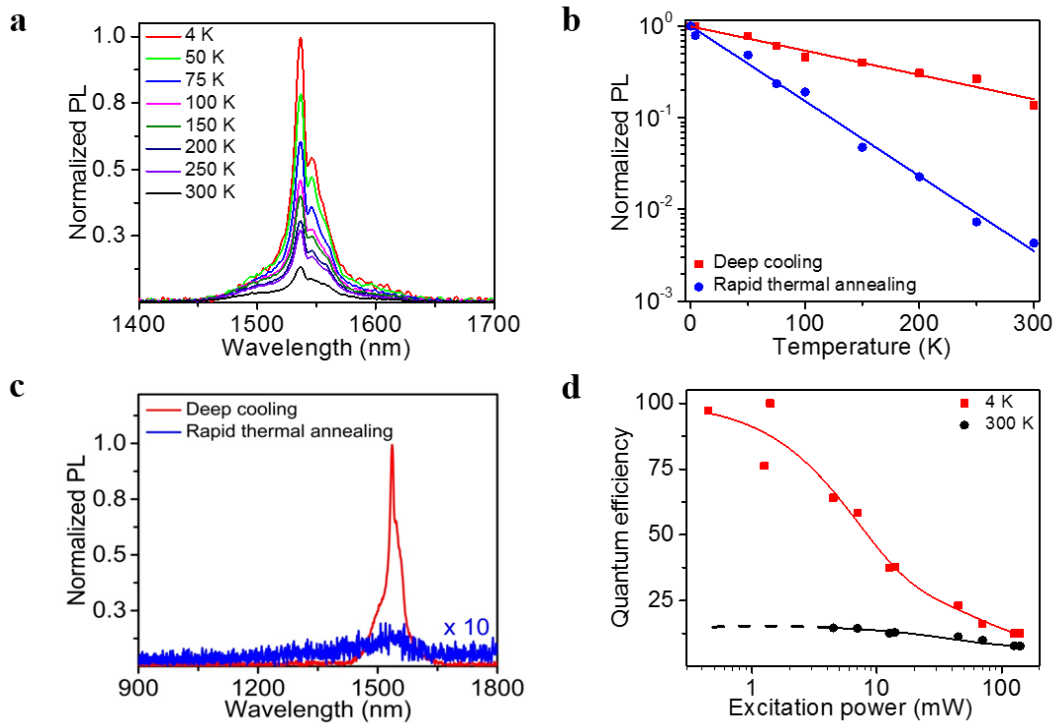

**Figure 1 │ Temperature-dependent PL spectra of the Er/O-Si samples. a**, PL spectra of the deep cooling (DC) processed samples measured at different temperatures. **b**, Peak intensity as a function of temperature for the DC- (red squares) and RTA- processed (blue dots) samples, respectively. **c**, Ambient room-temperature PL spectrum for the DC- (red line) and RTA-processed (blue line) samples. **d**, Quantum efficiency of the DC-processed sample as a function of the excitation laser power at 300 K (black dots) and 4 K (red squares), respectively.

The intrinsic single-crystalline silicon wafer was first doped with erbium and oxygen by ion implantation. The Er/O-implanted silicon wafer was cut into two sets of samples that separately went through the DC and RTA procedure (see Methods). The secondary ion mass spectrometric (SIMS) measurements in Supplementary Fig. 1a show that the Er and O dopants are mostly located at ~80 nm below the surface with a maximum concentration of $2.8 \times 10^{20}$ cm$^{-3}$ and $1.4 \times 10^{21}$ cm$^{-3}$, respectively. Fig.1a shows the temperature-dependent photoluminescence (PL) of the DC-processed samples. The sharp luminescence at 1.54 μm with a narrow full width at half maximum (FWHM, ~25 nm) represents the characteristic 4f transition of $Er^{3+}$.[5-10] The temperature-dependent PL of the RTA-processed samples is shown in Supplementary Fig. 2a. For better comparison, the normalized peak amplitude as a function of temperature for these two sets of samples is plotted in Fig. 1b. As the temperature increases from 4 K to 300 K, the radiative emission is thermally quenched by nearly 3 orders of magnitude for the samples that were treated by the standard RTA method. In contrast, the radiative emission is only quenched by less than 1 order of magnitude for the samples that went through the DC process. This results in a two-orders-of-magnitude increase of PL at room temperature for the DC-processed samples in comparison with the RTA-processed ones, as shown in Fig. 1c.

It is known that the photoexcited electron-hole pairs in silicon can recombine through three competing paths. The first path is radiative recombination and the other two paths are the nonradiative Shockley-Read-Hall (SRH) and Auger recombination.[19] The SRH recombination can be suppressed by lowering the operating temperature. The Auger recombination is highly dependent on the concentration of photogenerated electron-hole pairs. As the excitation laser power is reduced, the Auger recombination will be minimized. In Fig. 1d, we recorded the relative quantum efficiency (QE) of the DC-processed sample at 4 K to suppress the SRH recombination (PL shown in the Supplementary Fig. 2b). When the excitation laser power is reduced to < 1 mW, the Auger recombination is minimized and the excited electron-hole pairs can only recombine radiatively via the 4$f$ transition of the Er ions. In this case, the internal quantum efficiency (IQE) must saturate to 100% (red squares in Fig. 1d).[20] At 300 K, the IQE also saturates as the excitation laser power is reduced (PL shown in the Supplementary Fig. 2c). But the saturated IQE is ~7 times smaller than the value at 4 K. It indicates that the IQE of our DC-processed samples can reach a record of ~14% at room temperature, approximately 2 orders of magnitude higher than the samples treated by the standard RTA process. It is expected that the IQE will be further improved by tuning the Er and O concentration and their ratio.

Since the Er/O impurities are donor-type dopants in silicon, a PN junction diode can be formed simply by doping the sample with boron via ion implantation. The peak concentration of boron dopants is ~$10^{19}$ cm$^{-3}$ and the peak center is located at ~240 nm below the surface, approximately ~160 nm below the Er/O distribution peak (see Supplementary Fig. 1b). The DC process was applied to activate the Er/O and B dopants at the same time. A vertical PN junction diode was formed when a metal electrode was in contact with the underneath p-type silicon layer after the Er/O doped silicon layer is removed. The three-dimensional schematic and the cross-section of the device are depicted in Fig. 2a and b. An optical microscopic top view of the device is shown in the inset of Fig. 2c. The diode exhibits a rectifying current density *v.s.* voltage (*J-V*) curve

(Fig. 2c). When it is forward biased, the diode has a strong emission at room temperature with the spectral characteristics of Er ions (Fig. 2d). As the injecting current ramps up, the emission peak at 1.54 μm increases rapidly. A near-infrared optical microscope (0.9 - 1.7 μm) was employed to record the device lightening-up process (see the setup in Supplementary Fig. 3 and the video in the Supplementary Video). The emission images of the device at the driving current of ~35 mA, ~50 mA and ~65 mA are shown in Fig. 2e, f and g, respectively.

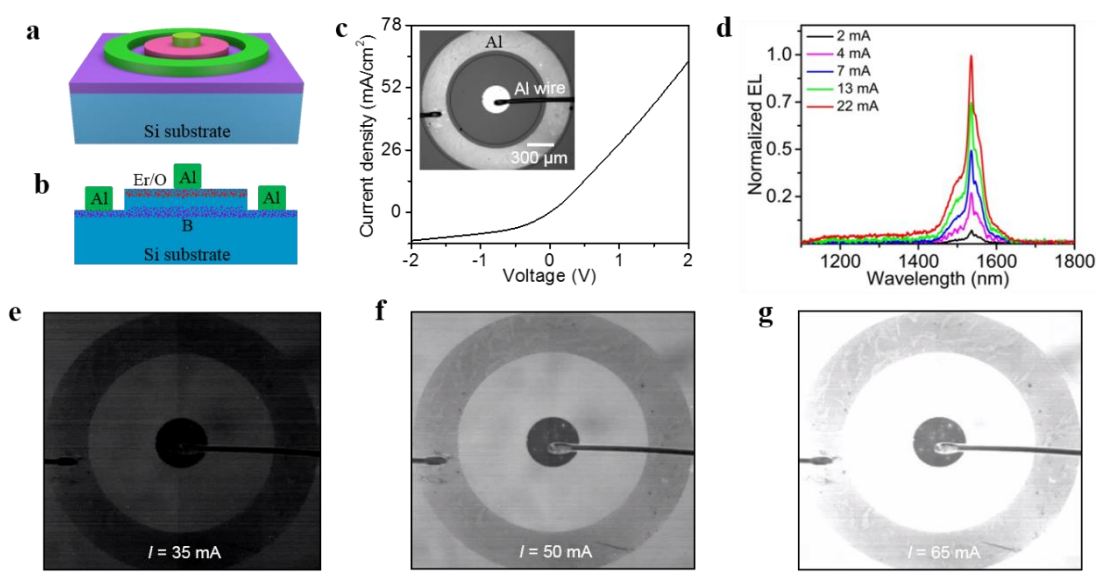

**Figure 2 | Optoelectronic properties of Er/O-Si LED devices treated by the deep cooling (DC) process.** Three-dimensional schematic (**a**) and cross-sectional illustration (**b**) of the device. **c**, *J-V* curve and optical microscope image (inset) of the device. Room-temperature EL spectra (**d**) and near infrared microscopic images (**e-g**) of the device emission under electrical pumping.

To understand the mechanism of the strong emission from the DC-processed sample, we used the high resolution transmission electron microscopy (HR-TEM) to examine the microstructure of the Er/O co-doped layer as shown in Fig. 3. It was previously reported that Er ions tend to aggregate during the slow cooling process of the standard RTA treatment.[17, 18] The aggregation of Er ions forms efficient nonradiative clusters that strongly suppress the emission of Er ions at room temperature.[5-10] The dramatic cooling of the DC process can freeze the dispersed Er dopants at high temperature (950 ℃) in an ultrashort time (possibly micro seconds), which helps prohibit the aggregation of Er ions. The cross-sectional TEM images show that Er nanocrystals as large as ~5 nm in diameter are clearly visible in the RTA-treated samples (Fig. 3a). The nanocrystals are more interconnected and aggregated. In the DC processed sample, Er nanocrystals are clearly smaller (< 1 nm) and more sparsely distributed as shown in Fig. 3b. The TEM image of a DC-processed intrinsic Si sample is shown in Fig. 3c for reference. This contrast is consistent with our expectation that the aggregation of Er ions is suppressed by the DC process, as a result of which the emission efficiency of Er/O co-doped Si samples is significantly increased at room temperature. The side effect of the DC process is that the thermal shock may create dislocations and cracks in the wafer surface.[21] The X-ray diffraction (XRD) pattern (Supplementary Fig. 4) shows that the silicon wafer remains a nearly perfect crystalline structure after the wafer is treated with the DC process. No cracks were observed by visual inspection. Indeed, TEM images show that some dislocations were formed in the silicon substrate 150 nm below the surface (Supplementary Fig. 5). The impact of these dislocations on CMOS transistors and the Si LED will be further evaluated. If needed, new measures, for example, a long time of thermal treatment at relatively low temperature may be able to remove these dislocations.

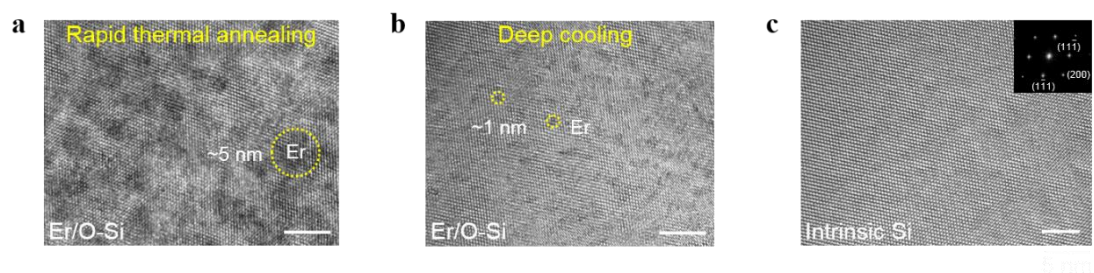

**Figure 3 │ Microstructure characterization of the fabricated Er/O-Si samples.** Cross-sectional HRTEM micrographs of (**a**) RTA- and (**b**) DC-processed Er/O-Si samples. The Er/O nanocrystal sizes distributed in a silicon matrix after RTA and DC treatments are approximately 5 nm (**a**) and 1 nm (**b**), respectively. **c**, Cross-sectional HRTEM micrograph of intrinsic single-crystalline silicon sample by the DC treatment was also manifest for comparison. Inset: Selected area electron diffraction (SAED) pattern of the intrinsic Si sample.

In short, we presented a DC process to treat the Er/O co-doped single crystalline silicon wafers. After the treatment, the aggregation of Er ions is suppressed, as a result of which strong room-temperature photoluminescence at telecommunication wavelength was observed from the silicon samples. We further doped the samples with boron dopants and eventually realized an efficient room-temperature silicon LED device at 1.54 μm. Given the fact that these results were realized on single-crystalline silicon, it is promising to develop room-temperature silicon lasers for fully integrated silicon photonics in future.[22]

## METHODS

**Preparation and characterization.** FZ intrinsic Si (100) wafers (Resistivity: ≥ 10 kΩ cm;

Thickness: 500 ± 20 μm) were purchased from Suzhou Resemi Semiconductor Co., Ltd, China. The silicon wafers were first cleaned with ethanol, deionized water and then immersed in Piranha solution (Sulfuric acid : 30% Hydrogen peroxide = 3 : 1) for 30 min at 90 ℃, followed by 2.5% HF etching. Erbium and Oxygen ions were implanted at the Institute of Semiconductors, Chinese Academy of Sciences in Beijing, China. The injection energy and dose of Er and O ions were 200 keV, $4 \times 10^{15}$ cm$^{-2}$, and 32 keV and $10^{16}$ cm$^{-2}$, respectively. After the implantation, the Er/O-implanted silicon samples went through the aforementioned cleaning procedure again before the deposition of a 200-nm-thick $SiO_2$ films by reactive magnetron sputtering (Delton multi-target magnetic control sputtering system, AEMD, SJTU). The deep cooling and standard RTA processes were applied to two Er/O co-doped silicon samples separately. After the capping $SiO_2$ film was removed in buffered oxide etchant (BOE), PL characterizations were firstly performed.

As shown in Supplementary Fig. 1a, SIMS technique (Evans Analytical Group, NJ, USA) was subsequently utilized to characterize Er and O dopant distribution profiles at the top 300 nm of silicon substrate. For Er element, the SIMS data are the summation of three Er isotopes (166, 167, and 168). For O element, no other isotopes were detected. The bump near the surface (blue line) likely comes from the contamination of native silicon oxide. To form a PN junction, boron dopants were later implanted into the Si substrate (see the section of LED fabrication and imaging). The SIMS profile of boron dopants is shown in Supplementary Fig. 1b.

**Deep cooling and RTA processes.** The deep cooling process was performed on the Er/O-Si samples in an upgraded dilatometer (DIL 805A, TA Instruments) in which the samples were annealed at 950 ℃ for 5 min by means of copper coil-based electromagnetic heating, followed by a flush of high purity He (99.999%) gas cooled in liquid $N_2$ (77 K). The RTA procedure for Er/O-Si samples was firstly conducted in a rapid thermal processing furnace (RTP-300) at 1050 ℃ for 5 min. After the high temperature treatment, the silicon samples were gradually cooled to room temperature with a high purity $N_2$ gas.

**PL measurements.** The PL measurements were performed using a Fourier transform infrared (FTIR) spectrometer (Vertex 80V, Bruker) equipped with a liquid-nitrogen-cooled Ge detector. A CW diode laser of λ = 405 nm (MLL-III-405, CNI, Changchun, China) providing a maximum excitation power of ~150 mW was employed as the excitation source. During the measurements, the Er/O-Si samples were mounted on the cold head of a Helium closed cycle cryostat (DE-202S, ARS), which allows for temperature adjustment from 4 K to ambient. (Fig. 1a and Supplementary Fig. 2a). For the excitation power dependent PL measurements, different optical filters including Notch filters, neutral density filters, and long-pass filters were utilized for avoiding the influence of the excitation source. The results are shown in Supplementary Fig. 2b and c.

**LED fabrication and imaging.** To form a PN junction on the Er/O co-doped silicon wafers, boron ions were implanted (80 keV and $5.2 \times 10^{14}$ cm$^{-2}$) into the silicon wafer with a peak concentration located ~240 nm below the surface (Supplementary Fig. 1b). All dopants were simultaneously activated in the subsequent deep cooling process. Since the Er/O dopants (donors) are located in a shallower layer, a vertical PN junction is formed between the n-type Er/O layer and the p-type B doping layer. A pair of co-axial electrodes was formed by UV photolithography (MDA-400M, MIDAS), deep silicon etching (SPTS ICP), and metal film deposition (Nexdep, Angstrom Engineering Inc.). As shown in Fig.2a and b, the internal electrode is in contact with n-type Er/O doping region and the external electrode in contact with the p-type region after the top Er/O layer is removed by deep silicon etching. All the microfabrication processes were performed at the Center for Advanced Electronic Materials and Devices, Shanghai Jiao Tong University.

After Al metal wire bonding (7476D, West Bond), the devices were integrated on a PCB board. The *J-V* curves of the Er/O-Si LED devices were measured using a digital sourcemeter (Keithley 2400) controlled by a Labview script. For EL measurements, the devices were electrically pumped by the sourcemeter and the emission spectra were collected using the same FTIR system for PL measurements.

Near Infrared imaging of the emission from the Er/O-Si LED devices was obtained at room temperature under optical microscope (BX53M, Olympus) equipped with a NIR camera (C12471-03, Hamamatsu). The setup is shown in the Supplementary Fig. 3. The recorded photographs and videos are shown in Fig. 2e, f, and g and Supplementary Video.

**Microstructure characterization.** The microstructure of Er/O-Si samples were extensively investigated with TEM and XRD techniques. XRD patterns were acquired with a poly-functional X-ray diffractometer (D8 advance, Bruker) shown in Supplementary Fig. 4. In comparison, XRD patterns of the intrinsic and Er/O-implanted Si samples without any thermal treatment were also collected. Talos F200X field-emission TEM was employed to take the EDX and TEM images (Supplementary Fig. 5) of a sample slide that was cut from the samples using focused ion beam

(FIB, GAIA3 GMU Model 2016). For EDX mapping measurements, the acceleration voltage was 200 kV, and other parameters were set as default. High resolution TEM images were further performed with JEOL JEM-ARM200F, as shown in Fig. 3.

**Data availability.** The data that support the findings of this study are available from the authors on reasonable requests, see author contribution for specific data sets.

**Acknowledgements** This work was supported by the National Science Foundation of China (61376001, 21703140 and 61874072), the China Postdoctoral Science Foundation (2016M601582), and the special-key project of the "Innovative Research Plan", Shanghai Municipality Bureau of Education. We thank Jianming Li (Institute of Semiconductors, CAS) for the ion implantation processing and Yihua Chen (Shanghai Jiao Tong Univ.) for the deep cooling treatment. We also thank Xuecheng Fu (Shanghai Jiao Tong Univ.) for the $SiO_2$ film deposition, Weiwei Wang (Spark Electro-Optics Co., Ltd.) and Lin Zhu (Cinv Co., Ltd) for the support in quantum efficiency and NIR imaging measurements.

**Author Contributions** Y.D. conceived the deep cooling idea and designed the experiments. H.W., J.J.H., J.H., and F.Y. conducted the experiments. H.W. and Y.D. analysed the data and wrote the manuscript. All authors reviewed and commented on the manuscript.

**Competing interests**

The authors declare no competing interests.

**Corresponding authors**

Correspondence and requests for materials should be addressed to Yaping Dan (yaping.dan@sjtu.edu.cn) and Fangyu Ye (fyyue@ee.ecnu.edu.cn).

**Author Information** Reprints and permissions information is available at www.nature.com/reprints. The authors declare no competing financial interests. Readers are welcome to comment on the online version of the paper. Publisher's note: Springer Nature remains neutral with regard to jurisdictional claims in published maps and institutional affiliations.

# Supplementary Figures

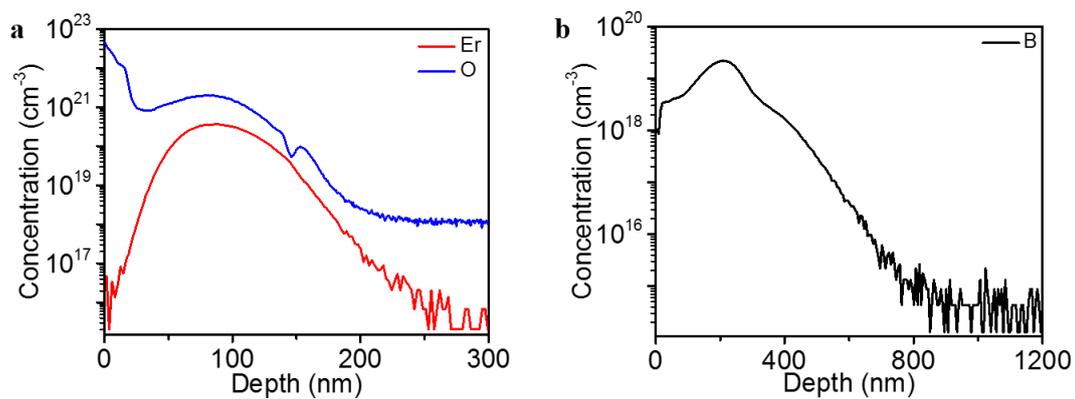

**Supplementary Figure 1 | SIMS characterizations of Er/O-Si samples. a**, Erbium (red line) and Oxygen (blue line) ion distribution profiles of the DC-processed Er/O-Si samples (to 300 nm below surface) by SIMS measurements. **b**, Boron ion (black line) distribution profile of Er/O-Si LED devices.

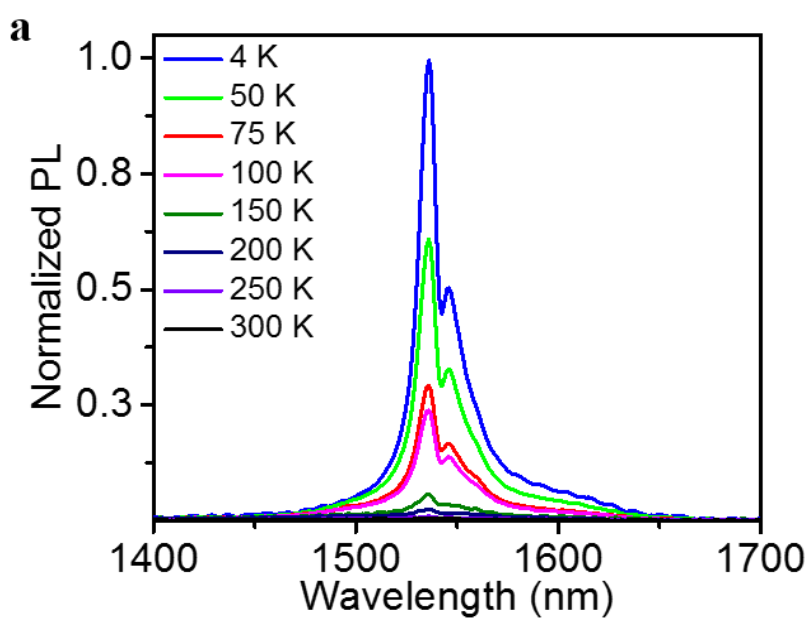

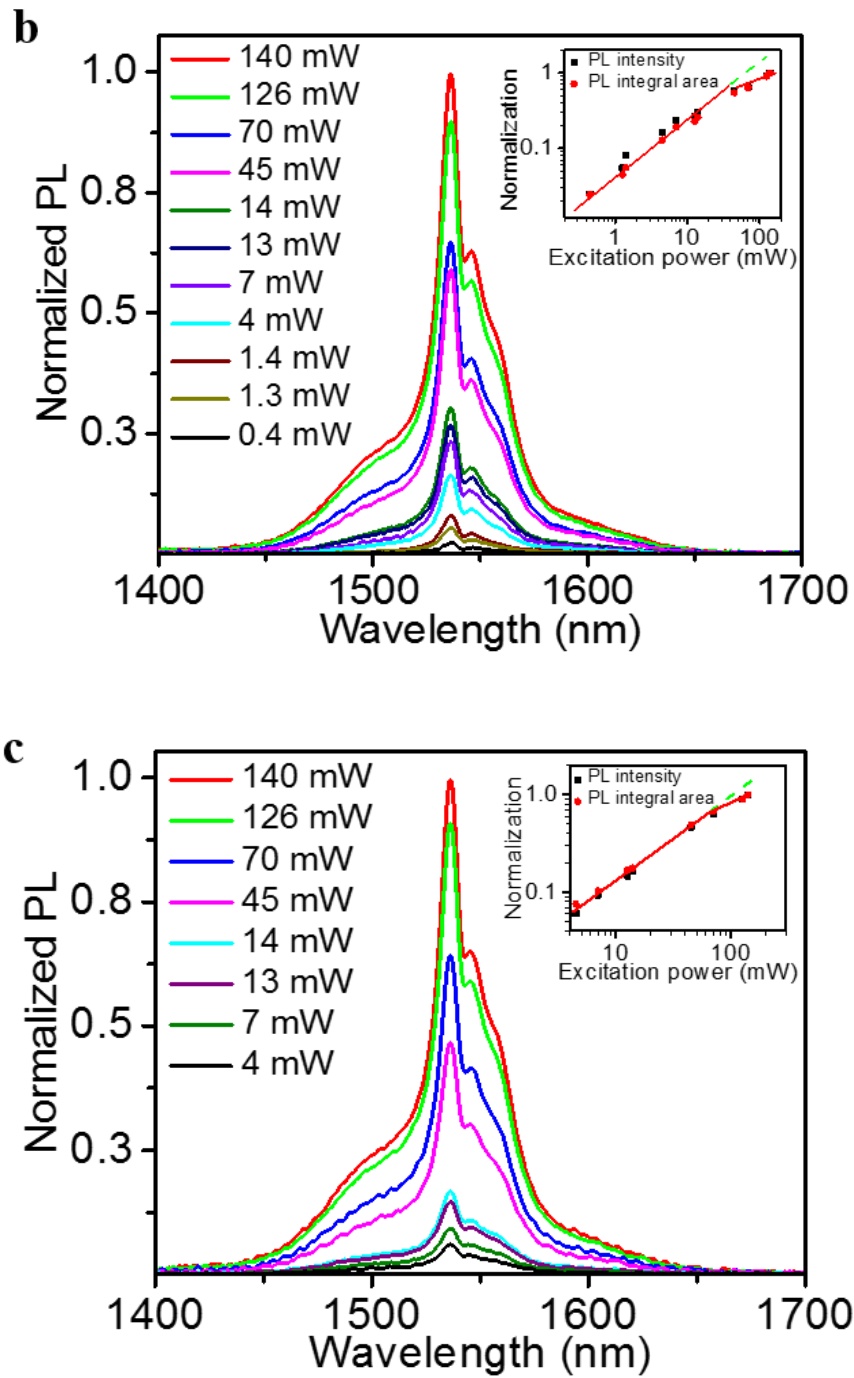

**Supplementary Figure 2 | Temperature- and excitation power-dependent PL spectra of Er/O-Si samples. a**, Temperature-dependent PL spectra of the RTA-processed samples, which were measured as a comparison with the results in Fig. 1a. **b**, Cryogenic- (4 K), and **c**, room-temperature (300 K) PL spectra as a function of excitation power for the DC-processed samples. The insets show the corresponding relationships between peak intensity (black squares) or integral area (red dots) and the excitation power, derived from **b** and **c**, respectively.

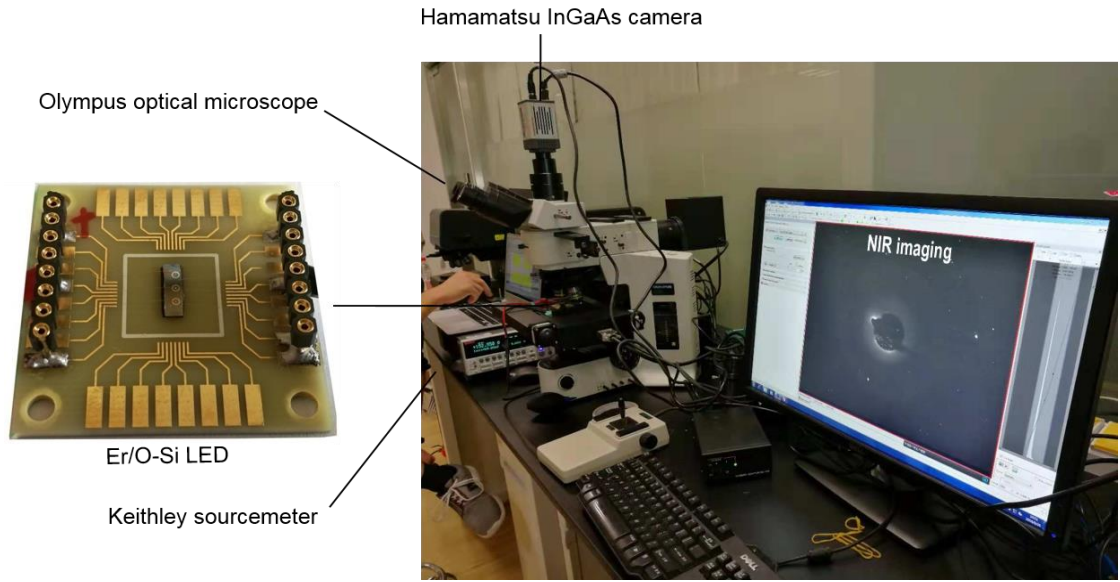

**Supplementary Figure 3 | Room-temperature NIR imaging setup.** A NIR Olympus optical microscope equipped with a Hamamatsu InGaAs camera (0.9 - 1.7 μm) was used to image the electrically-pumped emission (Keithley sourcemeter) from the Er/O-Si LED device that is treated with the deep cooling process. The optical image of PCB-integrated Er/O-Si LED devices was also displayed.

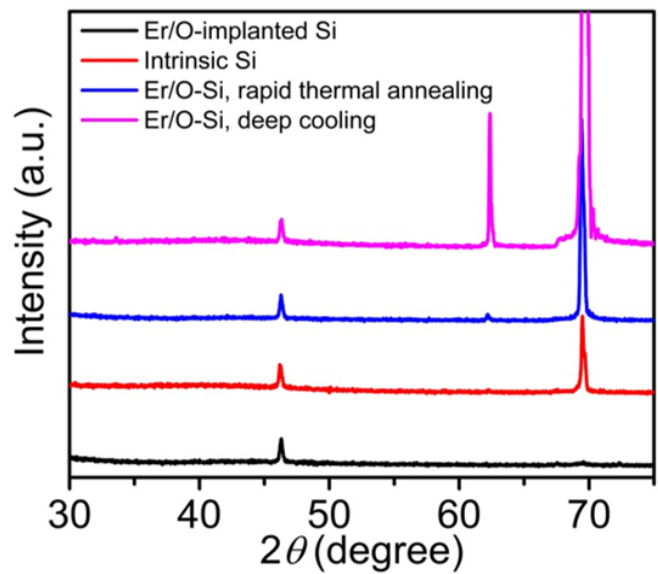

**Supplementary Figure 4 | XRD characterization.** XRD patterns of DC- (pink line) and RTA-processed (blue line) Er/O-Si samples. As comparison, the XRD results of intrinsic (red line) and Er/O-implanted (black line) silicon samples were also included in this figure.

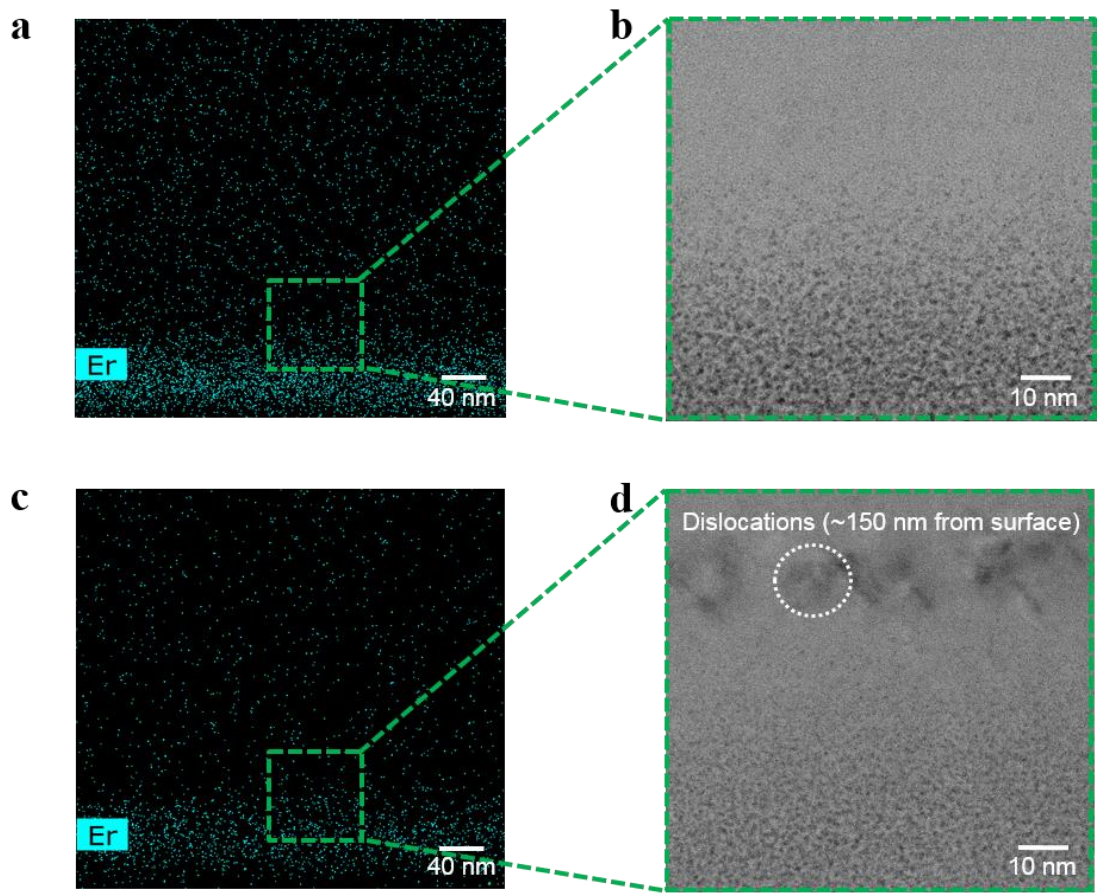

**Supplementary Figure 5 | TEM characterization. a**, Er EDX mapping, and **b**, cross-sectional TEM image of the RTA-processed samples. **c**, Er EDX mapping, and **d**, cross-sectional TEM image of the DC-processed samples.